\documentclass[%
 reprint,
 amsmath,amssymb,
 aps,
floatfix,
]{revtex4-2}

\UseRawInputEncoding

\usepackage{graphicx}% Include figure files
\usepackage{dcolumn}% Align table columns on decimal point
\usepackage{bm}% bold math
\usepackage{float}
\usepackage{siunitx}

\newcommand{\etal}{\textit{et al. }}
\newcommand{\m}{$\mu$m~}

\begin{document}

\preprint{APS/123-QED}

\title{Frustrated “run and tumble” of swimming E-coli bacteria in nematic liquid crystals}% Force line breaks with \\

\author{%%%% Author details
Martyna Goral$^{1,2}$, Eric Cl\'{e}ment$^{1}$, Thierry Darnige$^{1}$, Teresa Lopez-Leon$^{2}$, Anke Lindner$^{1}$}

%%%%%%%%% Insert author address here
\address{$^{1}$Laboratoire de Physique et M\'{e}canique des Milieux H\'{e}t\'{e}rog\`{e}nes, UMR 7636, CNRS, ESPCI Paris-PSL, 75005 Paris, France\\
$^{2}$Laboratoire Gulliver, UMR 7083, CNRS, ESPCI Paris-PSL, 75005 Paris, France}

\date{\today}% It is always \today, today,

\begin{abstract}

In many situations bacteria move in complex environments, as for example in soils, oceans or the human gut-track microbiome.  In these natural environments, carrier fluids such as mucus or reproductive fluids show complex structure associated with non-Newtonian rheology. Many fundamental questions concerning the the ability to navigate in such environments remain unsolved due to the inherent complexity of the natural surroundings. Recently, the interaction of swimming bacteria with nematic liquid crystals has attracted lot of attention. In these structured fluids, the kinetics of bacterial motion is constrained by the orientational molecular order of the liquid crystal (or director field) and novel spatio-temporal patterns arise from this orientational constraint, as well as from the interactions with topological defects. A question unaddressed so far is how bacteria are able to change swimming direction in such an environment. In this work, we study the swimming mechanism of a single bacterium, \textit{E. coli}, constrained to move along the director field of a lyotropic chromonic liquid crystal (LCLC) that is confined to a planar cell. In such an environment, the spontaneous “run and tumble” motion of the bacterium gets frustrated: the elasticity of the liquid crystal prevents flagella from unbundling. Interestingly, in order to change direction, bacteria execute a reversal motion along the director field, driven by the relocation of a single flagellum to the other side of the bacterial body, coined as a "frustrated tumble". We present a detailed experimental characterization of this phenomenon, exploiting exceptional spatial and temporal resolution of bacteria and flagella dynamics during swimming, obtained using a two color Lagrangian tracking technique. We suggest a possible mechanism behind the frustrated run and tumble motion, accounting for these observations.

\end{abstract}

\maketitle

%\tableofcontents

\section{Introduction}
%\protect\\ The line break was forced \lowercase{via} \textbackslash\textbackslash

Microorganisms, such as bacteria, are not only of great importance in bio-medicine or food industry, but they also constitute very interesting active fluids \cite{J.Schwarz-LinekJ.ArltA.JepsonA.DawsonT.VissersD.MiroliT.PilizotaV.A.Martinez2016}. Exotic phenomena emerge from the interplay between bacterial dynamics and the properties of the fluid in which they swim; for example, the suspension of bacteria into a Newtonian fluid leads to surprising non-Newtonian properties and in particular a decrease of the effective viscosity below the viscosity of the suspending fluid \cite{Gachelin2014,Lopez2015}, while the fluid in turn can induce complex swimming behaviors in the bacterial suspension, as enhanced swimming velocities in colloidal suspensions \cite{Kamdar2022TheMotility} or straightening of bacteria trajectories in polymer solutions \cite{Patteson2015d}.

In Newtonian aqueous media, model peritrichous bacteria, as \textit{E. coli}, follow trajectories that resemble a random walk with no specific spatial orientation. 
This random walk results from the so called "run and tumble" dynamics described in a seminal work by Berg \etal \cite{BergHowardCBrown1972,Darnton2007a}. 

Ideally, it consists of a run phase, where the bacterium swims in straight line, that is periodically interrupted by a tumble phase, where the swimming direction changes. In the run phase, the bacterium is propelled by a bundle of flagella that turns counter-clockwise (CCW). The tumble is triggered when at least one of the flagella starts turning clockwise (CW), provoking the opening of the bundle. 

The reversal in rotation direction and the associated torques acting on the flagellum leaving the bundle have been shown to lead to polymorphic transformations, where the flagella geometry as well as the handedness can change \cite{Turner2000,Darnton2007,Vogel2013}. A typical polymorphic transformation concerns the change from a "normal" left-handed flagellum as those comprised in the bundle to a "curly" form of opposite handedness as in the single CW rotating flagellum. In order to keep the total torque equal to zero, the body responds to the consequent opening of the bundle by changing its orientation. Eventually, the flagellum rotation switches back to CCW as well as to a "normal" configuration, leading to the reorganisation of the bundle and to a new run phase. The frequency of tumbles is controlled by a chemotaxis signaling network, enabling navigation toward or away from certain regions in the medium \cite{Berg+2018,PhysRevE.48.2553}.

The swimming pattern of peritrichous bacteria have been intensively studied in Newtonian aqueous media. Yet bacterial natural environments are usually complex and non-Newtonian; examples range from crystalline mucus to structured soils \cite{Viney1993,Suarez2003,Bansil2013,Costerton1995, Wilking2011, Denton1968,Kimsey1990, Jung2010,Bhattacharjee2019}. Many intriguing observations have been made in complex fluids as polymer solutions or colloidal suspensions and include enhanced bacterial mobility \cite{Patteson2015d, Kamdar2022TheMotility}, straightening of bacteria trajectories \cite{Patteson2015d, Kamdar2022TheMotility}, suppression of tumbles \cite{Patteson2015d} or even the appearance of new swimming patterns as slow random walks \cite{Qu2018}. They are all attributed to the complex interactions of the motile bacteria with the complex fluid, either through the effective fluid properties or the fluid inhomogeneities on the scale of bacterial body or flagella \cite{Martinez2014}.

Recently, the investigation of bacteria motility has been extended to fluids presenting a structure at the molecular or supra-molecular scale, as for example liquid crystals  (LCs) \cite{Kumar2013,Mushenheim2014a,Aranson2018,Chi2020,Daddi-Moussa-Ider2018,Rajabi2021,Krieger2015}. LCs are anisotropic viscoelastic materials displaying long range partial orientational and positional order. Their physico-chemical properties have been well established over decades of intensive research. Liquid crystals are responsive to external conditions -- including electro-magnetic fields, substrates and interfaces, temperature and concentration gradients, or pH -- what makes them adaptable and controllable. Besides, they are often used as models for biological systems, providing conditions close to those of natural bacterial environments \cite{Rey2010}.

Recent studies on bacterial motion in LCs have shown that the swimming patterns resembling an isotropic random walk, originating from "run and tumble" dynamics, typically observed in Newtonian aqueous solutions is suppressed by the elasticity of LCs. Instead, the bacteria are constrained to follow the direction of molecular alignment of the LC or director field \cite{Kumar2013,Mushenheim2014a,Zhou2014,Peng2016a}. Intricate collective dynamics result from elastic and hydrodynamic interactions between bacteria and topological defects, opening very interesting perspectives for future research. However, it remains unclear how changes in swimming direction occur in these anisotropic media and how such changes are linked to the "run and tumble" mechanism. 

In this paper, we investigate the swimming mechanism of \textit{E. coli} bacteria dispersed in a lyotropic chromonic liquid crystal (LCLC) that is uniformly aligned in a planar cell. We observe linear trajectories aligned with the nematic field, consistently with previous observations. Interestingly, we observe recurring re-orientation episodes, where the bacteria change their swimming orientation while maintaining their direction aligned with the director field. These "reversals" entail the relocation of at least one flagellum on the other side of the bacterial body. We show that this configurational change is in most cases associated with a polymorphic transformation in the flagellum and that it occurs at a frequency similar to that of tumbling. We propose a possible mechanism for bacteria locomotion in the reverse mode. Our results suggest that this novel reversal swimming pattern emerges from a frustrated "run and tumble" resulting from the elastic constrains imposed by the liquid crystal.

\section{Materials and Methods}
%\protect\\ The line break was forced \lowercase{via} \textbackslash\textbackslash

\subsection{Bacterial strains and culture}

We use a wild type strain of \textit{E. coli} (ABAD62, derived from K12, see more details in reference \cite{Junot2021} Supplemental Material)  that allows us to visualize the body as well as the bacteria flagella. It has an encoded GFP production for the bacterial body and special flagella receptors for fluorescent dye and can thus be used to stain bacteria with two different colors. 
Bacteria were first inoculated in a growth culture medium LB with ampicillin antibiotic over night at 30° C and 250 RPM. The following morning 10\% of the bacterial solution was suspended in TB growth medium for 4-5 hours, until it reached an optical density of 0.5. 
The bacteria were centrifuged at 5000 RPM for 6 minutes twice in Berg's motility buffer (BMB) and rinsed with BMB in between. 
At this point Alexa dye (Maléimide Alexa Fluor™ 647 C2 purchased from ThermoFisher scientific) was added to the bacterial solution and the sample was put on a rotation plate for 2 to 3 hours. Finally the bacteria were washed (centrifuged and re-suspended in motility buffer) three times to remove the excess of dye, re-suspended and ready to use. The final bacterial suspension has an OD of 15.

\subsection{Solution preparation}

The \textbf{liquid crystal} chosen was disodium 5,5'-[(2-hydroxy-1,3-propanediyl) bis(oxy)] bis[4-oxy-4H-1- benzopyran-2-carboxylate] (from Sigma Aldrich), also known as DSCG, because of its compatibility with bacterial life. It is a lyotropic chromonic liquid crystal, whose properties result from the spontaneous stacking of disk-like molecules in water. Concentration or temperature can trigger phase transitions from the isotropic state, to the nematic, columnar, or crystalline state, as concentration increases or temperature decreases. Lyotropic chromonic liquid crystals are characterized by unusual values of their viscoelastic constants. In particular, they exhibit an anomalously low elastic constant associated to twist deformation, which sometimes results in spontaneous chiral symmetry breaking \cite{Lydon2010,Habibi2019}.

Our liquid crystalline suspensions were prepared by dispersing the DSCG in bacterial motility buffer. To help the dispersion of DSCG in the motility buffer, we used a vortex shaker and gentle heating. We established the phase diagram of DSCG in the motility buffer (see Appendix) in order to find the sweet spot where the nematic phase appears at conditions of temperature and viscosity compatible with the swimming of \textit{E. coli}. According to the phase diagram, the transition from an isotropic to a fully nematic phase can be obtained by decreasing temperature. This transition occurs at lower temperatures for smaller DSCG concentrations. Higher liquid crystal concentrations lead to higher viscosities where bacteria swimming is slowed down, whereas smaller temperatures also lead to a reduced bacterial mobility \cite{Maeda1976a}. A compromise has thus to be found between these two competing effects. To facilitate the experimental observations we decided to work at room temperature and the appropriate conditions thus necessitate to choose the lowest DSCG concentration resulting in a full nematic phase between 21°  C and 25°  C, corresponding to $\Phi$ =12 wt \%.  The DSCG viscosity could not be measured with a simple rheometer since it is anisotropic and time dependent as it has been shown by Habibi \etal \cite{Habibi2019}. They used micro-rheology to characterize this liquid crystal and measured viscosities in the direction parallel to the LC director. In the other directions the fluid has a complex and strongly time dependant behavior and they were not able to derive an associated effective viscosity from their measurements.

For a concentration of $\Phi$ =12 wt\%, the viscosity is $\mu$ = 18.5 mPa.s at room temperature. Note that the viscosity increases very rapidly with concentration and for $\Phi$ =13.3 wt \%, $\mu$ = 63.5 mPa.s. It has been shown that the viscosity to immobilize flagellated bacteria like  \textit{E. coli} is around 60 mPa.s at room temperature \cite{Greenberg1977} and indeed in our experiments, \textit{E. coli} ceased to be motile from $\Phi$ =13 wt \% onward, which is in agreement with the corresponding viscosity. 

To prepare a solution of liquid crystals containing bacteria, 5\% of the bacterial solution is added to the liquid crystal. The initial liquid crystal concentration has to be 12.63 wt\% in order to obtain final concentrations of 12 wt\%. Bacteria concentrations are keep small enough to be able to observe one bacteria at a time and to avoid any interaction between bacteria. 

For the \textbf{Newtonian viscous fluid} we used Ficoll 400 (F8016 reference), purchased from Sigma-Aldrich, taking the form of a powder that we dispersed in motility buffer. The presence of small polymer fragments has been shown to increase bacteria mobility as it can be metabolized by the latter \cite{Qu2018, Martinez2014}, preventing a clear link between the viscosity of the surrounding fluid and the bacteria swimming speed to be established. We thus assured that small polymer fragments were removed by performing a dialysis in the laboratory on the Ficoll 400 solution, during a week.
Finally, experiments were performed in dialyzed Ficoll solutions, with measured constant viscosity of $\mu_{Ficoll400-dialyzed} = 18\ \si{mPa.s}$ at room temperature. The viscosities were measured with an Anton Paar MCR 501 rheometer, with a conical geometry CP60-0.5.

\subsection{Sample preparation and bacteria tracking}

The observation chamber was formed by either two glass slides or a glass slide and a cover slip. The glass slides were well controlled diamond-scratched glass surfaces \cite{Suh2018} enabling perfect LC alignment.
 The cover slips on the other hand were rubbed manually on velvet to create micro-scratches, which enabled a better image resolution compared to the glass slides. Observation through the cover slips was possible using a water objective with small working distance (C-Apochromat 63×/1.2W). 
The glass surfaces were separated by adding PS beads (20 \m) to the liquid crystal bacteria solution. The samples were sealed to prevent external flows and evaporation, with either epoxy glue or nail polish, both bio-compatible.
Bacteria were observed with an inverted microscope (Zeiss-Observer, Z1), Colibri LED illumination, a camera (Hamamatsu Orca-flash 4.0), equipped with a beam splitter and a 3D tracking device \cite{Darnige2017a,Junot2021}. Using the 3D tracking we were able to record bacteria trajectories over long times while having access to recorded images of the bacterial body and flagella. 
For data analysis, custom Matlab scripts were written to detect bacteria reversals.

%%%%%%%%%%%%%%%%%%%%%%%%%%%%%%%%%%%%%%%%%%%%%%%%%%%%%%%%%%%%%%%%%%%%%

\section{Results}

Experimental observations were performed tracking individual \textit{E. coli} bacteria swimming either in the Newtonian solution of  Ficoll 400 or in the DSCG liquid crystal. At the chosen experimental conditions of temperature and concentration, the DSCG has the same viscosity as Ficoll 400, that is, $18\ \si{mPa.s}$. The bacteria are confined to a 20 micron-high flat cell with planar LC alignment so that their trajectories are considered confined to a nearly two dimensional plane. Bacterial body and flagella are colored fluorescently and can be observed simultaneously during tracking. 16 tracks have been recorded in Ficoll and 64 in the LC of a typical duration of $ 40 \pm 37\ \si{s}$. 

\subsection{Reversal kinematics : \protect\\ experimental observations}

We will first describe typical swimming trajectories observed in the isotropic viscous fluid and in the liquid crystal, focusing on trajectories with a change in swimming direction.  Fig. \ref{fig:trajectoire_vitesse_DSCG_Ficoll}(a) shows snapshots of a bacterium describing a typical trajectory in the Ficoll viscous solution. From these snapshots, separated by a time interval of 0.5 s, it can be seen very clearly how the flagella bundle opens up and several individual flagella become visible, corresponding to a tumble event. At the end, the bundle is fully formed again. It is also clear that the bacterium orientation changes during the tumble, in agreement with the classical picture of run and tumble motion. Fig. \ref{fig:trajectoire_vitesse_DSCG_Ficoll}(b) shows a typical 3D-trajectory indicating bacterial motion without a particular spatial orientation, as expected.

Figure \ref{fig:trajectoire_vitesse_DSCG_Ficoll}(c) shows snapshots of a bacterium moving in the liquid crystal. One remarks immediately that the bacterium moves along a single direction, corresponding to the direction of alignment of the LC director. Interestingly, the bacterium, that initially swims downwards, starts swimming upwards at some point: it changes its orientation by 180°, while keeping the same swimming direction. We will refer to this change in orientation as a "reversal". While the bacterium is swimming in the forward direction, it is pushed by a well-formed bundle. After some time, one observes a single flagellum appearing gradually on the other side of the bacterial body, a phenomenon that coincides with the reversal. We will refer to this flagellar configuration as "bi-polar". The single flagellum then disappears and another 180° reorientation is observed. The bacterium recovers its "normal" configuration with a single flagella bundle. We will discuss this flagellum relocation in more detail below, but from these observations it is already clear that the alignment between flagella and body is maintained throughout the reversal, in stark contrast with the bundle opening observed during a classical tumble in the isotropic medium. Note that such a reversal behavior, triggered by the relocation of a single flagellum to the other side of the bacterial body has been observed in most of our experiments, even if sometimes more complex behaviors were also detected (see Fig. \ref{fig:kink-deformation}~(e)). In a large majority of "bi-polar" flagella configurations, the bacterium is pushed by the single relocated flagellum and swims in the opposite direction compared to the "normal" configuration. A typical trajectory is shown in Figure \ref{fig:trajectoire_vitesse_DSCG_Ficoll}(c), which confirms the existence of a remarkable bacterial dynamics in the liquid crystal, showing that bacteria are able to reorient in this constraining environment, by moving back and forth along the director. 

We have analyzed the swimming velocities of the bacteria in both media, obtaining the velocity distributions shown in Figure \ref{fig:trajectoire_vitesse_DSCG_Ficoll}(c). The resulting mean velocity for a bacterium swimming in the viscous Newtonian fluid is $ 1.22 \pm 1.24\ \si{\mu m.s^{-1}}$, whereas we find $ 2.32 \pm 0.9\ \si{\mu m.s^{-1}}$ for the liquid crystal. In the LC, the histogram contains values from all possible swimming mechanisms (normal, bi-polar or more complex, rare configurations), explaining the large width of the velocity distribution. A more detailed analysis will be performed below (see Fig.\ref{fig:mechanism_crosssection}). These two average swimming velocities are comparable within the error bars, and are found to be significantly lower than the values reported for the same bacteria strain in a buffer solution, 26 $\si{\mu m .s^{-1}}$ \cite{Junot2021}. Our fluids are about 20 times more viscous than the buffer solution. According to Martinez \etal and Qu \etal, the swimming velocity is expected to be inversely proportional to the fluid viscosity under operating conditions where a constant motor torque is expected, as it is the case here \cite{Martinez2014, Qu2018}. Based on the experimental results by Martinez \etal \cite{Martinez2014} establishing an exact relationship between viscosity and swimming velocity, the typical velocity predicted for our {\it E-coli} bacteria, in a fluid of viscosity of 18mPas, is around $4 \si{\mu m .s^{-1}}$.  While this prediction correctly captures the order of magnitude of the swimming speed observed in our experiments, we note however that the velocities observed in our experiments are a least a factor two smaller. This is most likely due to the strong confinement induced by the experimental cell (height $20\si{\mu m}$), previously shown to lead to decreased swimming velocities \cite{Figueroa-Morales2015}. The velocity probability distribution functions also show that some bacteria are able to reach higher speeds in the LC than in the Newtonian viscous fluid. Similar observations on increased swimming speed in complex fluids have also been made in colloidal suspensions \cite{Kamdar2022TheMotility} or polymer solutions \cite{Patteson2015d}, consistently with our results.

In the following we will rationalize this novel and surprising mechanism of change in swimming direction, investigating first what triggers a reversal, then, how a flagellum can relocate from one side of the bacterial body to the other, and finally, how a single flagellum could be able to push the entire bacterium. 

\subsection{Reversals triggered by the initiation of a tumble}

The reversals observed in the LC solution occur at random positions within the sample and we have checked that they do not coincide with tactoid boundaries or any visible defects in the solution. Visual observation under crossed polarizers indeed indicates a uniform director field in the liquid crystal. From a mesoscopic point of view, the liquid crystal is composed of elongated aggregates, with an average length of 8 nm, which are too small to represent an obstacle for the bacteria \cite{Agra-Kooijman2014ColumnarCromoglycate}. We can thus exclude mechanisms such as those recently proposed for bacterial reversals observed in mucus \cite{Figueroa-Morales2019a}, Carbopol solutions \cite{HectorUrraPhD2021} or homeotropic liquid crystals \cite{Zhou2017b}. In these systems, where a spatial heterogeneity is present, bacterial reversals have been attributed to the encounter of "obstacles" along the bacteria swimming path. The reversals occur when the bacterial body is suddenly stopped and the compression on the flagellar bundle causes buckling, disassembly and reorganization on the other side of the bacterium \cite{Figueroa-Morales2019a}. In Zhou \etal \cite{Zhou2017b} the reversal occurs perpendicular to the nematic director, playing the role of the encountered "obstacle" by limiting displacement along the original swimming direction.

We hypothesize that the observed reversals in the LC are triggered by the initiation of a tumble motion, and thus one flagellar motor changing rotation direction to CW, causing a flagellum to leave the bundle and to relocate to the other side of the bacterial body. Upon switching back to CCW, the flagellum relocates back to its initial position and the bundle is formed again. Control experiments with "smooth swimmers", mutant \textit{E. coli} with genetically suppressed tumbling, showed no reorientation, supporting the idea that reversals are indeed linked to the initiation of a tumble. We will discuss next if the observed timescales for the reversals in the LC are compatible with this picture, taking advantage of the excellent resolution on the flagella dynamics of our observations. We will compare the observed timescales with those obtained in the Newtonian medium, where we know that classical run and tumble motion is preserved. We will also verify that our observations are compatible with previous observations in buffer solutions and equally viscous surroundings.

Two important time scales can be discussed, first the \textbf{frequency of tumbles} (corresponding in first approximation to the inverse of the run times), and second, the \textbf{tumble duration}. In our picture, the tumble frequency corresponds to the reversal frequency (or the inverse of the duration of swimming in a "normal" configuration) whereas the tumble duration would correspond to the time of swimming in the "bi-polar" configuration.  Berg \etal \cite{BergHowardCBrown1972,Darnton2007a} identified a typical run time of 1s for {\it E. coli} in buffer solution (corresponding to a tumbling frequency of about 1Hz) and a tumbling time of 0.1s. It has recently been shown that the run time distribution is very large, consistently with the large distribution of the  CCW  motor rotation duration and that a correct determination of the latter requires thus the analysis of a large number of long bacteria trajectories \cite{Figueroa-Morales2020, Korobkova2004FromBacterium}. The number of trajectories of limited length obtained in this work is thus not sufficient to get a clear picture on the run time distributions, but we can state that typical run times in Ficoll, as well as "normal" mode swimming times in the LC, are longer than 10s, a value significantly larger than that reported by Berg in the buffer solution. Qu \etal \cite{Qu2018} have stated that typical CW and CCW motor rotation intervals should be relatively insensitive to viscosity, as long as flagella rotation speeds remain not too low (typically above 50 Hz). This would indicate an unchanged tumbling frequency as a function of viscosity. However, their experiments have been performed for relatively low viscosities (below 5 mPas), and do thus not correspond to the regime investigated here.  Recent observations by Patterson \etal \cite{Patteson2015d} indicate that run times can be increased for high solvent viscosities where bacteria operate close to a stall regime (flagella rotation frequencies below 50Hz) leading to low swimming velocities and low flagella rotation rates, in agreement with our observations.

\begin{figure*}[h]
\centering
\includegraphics[width=\textwidth]{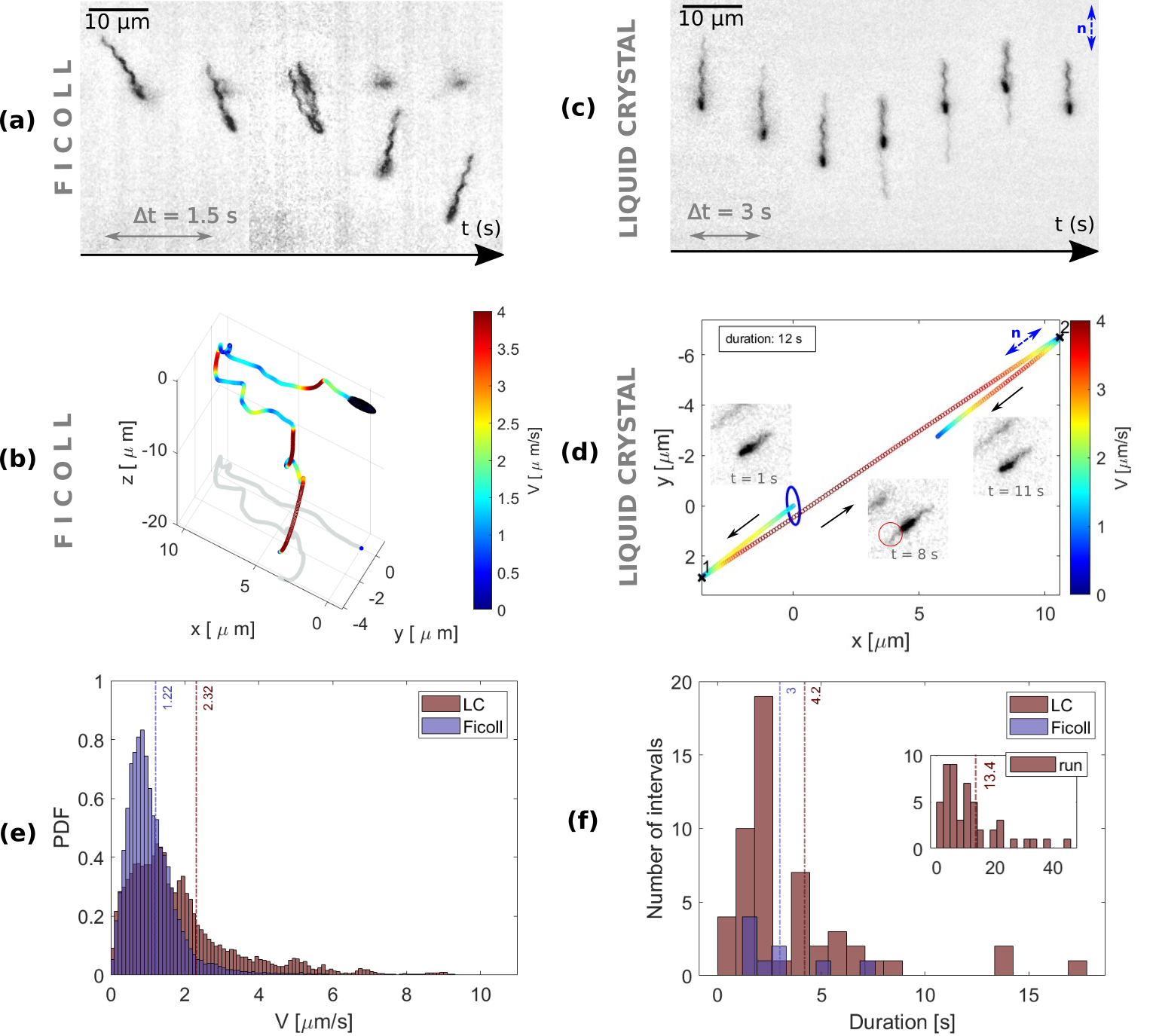}
\caption{(a) Snapshots of a free swimming \textit{E. coli} bacterium performing a run and tumble motion in a 3D space, in a Ficoll solution of viscosity $18 \si{mPa.s}$, at room temperature. The time duration between two snapshots is 1.5 s. Both the flagella and the bacterial body can be seen, during the run phase when the flagella form a bundle, and during the tumble, when the flagella separate. The final swimming direction is different from the initial one.  (b) Trajectory of a tracked bacterium changing its swimming direction in the Ficoll solution. The color scale corresponds to the absolute velocity values, and the blue ellipsoid represents the bacterial body when the track starts (only for scale purposes).(c) Snapshots of a bacterium swimming in a liquid crystal (DSCG 12 wt\%, $18 \si{mPa.s}$, room temperature). Confined in a $20 \si{\mu m }$ high observation chamber. The nematic director is along the vertical direction, as well as the bacterium alignment. The time duration between two snapshots is 3 s. The bacterium reverses its swimming direction successively from downwards to upwards. While the bacterium swims upwards, the bundle is split in two, in a bipolar configuration. The double arrow indicates the LC director (d) Trajectory of a bacterium reversing by 180 degrees its swimming direction in DSCG. The reversals are marked on the trajectory by crosses. The images show the body and flagella of the bacterium, corresponding to the different configurations: for t = 1 s and t = 11 s  normal configuration and for t = 8 s bipolar configuration. The relocated flagellum is pointed with a red ellipse and the arrows highlight the swimming direction. (e) Probability distribution function (PDF) histogram of absolute velocities for 16 tracked bacteria in Ficoll solution (red) and 64 tracked bacteria in DSCG liquid crystal (blue). The corresponding average velocities are $V = 1.22 \pm 1.24 \si{\mu m.s^{-1}}$ for Ficoll and $V = 2.32 \pm 0.9 \si{\mu m.s^{-1}}$ in the LC, represented by the dashed lines.(f) Histogram of tumble duration for bacteria swimming in Ficoll and LC, and the histogram for run duration in Ficoll.}
\label{fig:trajectoire_vitesse_DSCG_Ficoll}
\end{figure*}

Figure \ref{fig:trajectoire_vitesse_DSCG_Ficoll}(f) shows the time distributions for tumbling, in Ficoll, and swimming in the "bi-polar" mode, in the LC. Note that we refer to a tumble as the period over which the bacteria bundle is not fully formed, a definition that differs from the one used by Berg \etal \cite{Darnton2007a}, using a criterion based on the reduction of swimming speed in combination with bacterial reorientation. The trajectories in Ficoll have been analyzed by hand, inspecting the flagella bundle. As in the LC the normal and bi-polar swimming modes correspond to opposite swimming orientations (see Fig. \ref{fig:mechanism_crosssection}~(c)) automated detection was possible. The average values are respectively $t_{bi-polar,DSCG}=4,2s \pm 3.8 s$ and $t_{tumble,  Ficoll}=3s \pm 2s$, comparable to each other within the error bars. This is a first indication that the "bi-polar" swimming phase could indeed correspond to a tumble phase, as similar time scales are observed for both modes. 

It is less straight forward to compare these values to the values reported in the literature. Mainly, because of the different criteria used to establish the tumble time, which necessarily lead to different measured values. Indeed, in most investigations, the criteria used  to establish the tumble time were not based on the unbundling of the flagella, as clear observation of the flagella dynamics was lacking. Only the recent work by Junot \etal \cite{Junot2021} provided typical durations of flagella unbundling in a buffer solution, finding an average value of 0.8s. Qu \etal \cite{Qu2018} have argued that even if the duration of the CW rotation of the motor remains unchanged, the time to reform the bundle is strongly increased by the viscosity of the surrounding fluid. They claim that the rotation frequency of the flagella is reduced (as well as the swimming speed) due to the increased viscosity, and they hypothesize that 20 rotations are necessary to completely reform a bundle once the flagella has recovered its CCW rotation direction. In viscous surroundings, the tumbling time, defined as the time during which the bundle is not fully formed, is thus set by the time to reform the bundle, rather than by the time the flagella motor turns CW. We observe typical rotation frequencies between 5Hz and 10Hz (Fig. \ref{fig:mechanism_crosssection}~(e)) leading to a bundle formation time of typically 2-4s, in good agreement with our experimental observations. Note that the bundling time resulting from these fluid structure interactions dominates largely the time the flagella motor turns clockwise, so that possible increased durations of CW rotation in the high load regime \cite{Fahrner2003} possibly play a minor role.

Our analysis of the observed run and tumble times, as well as the time distributions of "normal" and "bi-polar" swimming, shows comparable values for the tumbling frequency in the Newtonian fluid and the reversal frequency in the liquid crystal, even if no thorough statistics are available for this variable. This suggests that reversals are triggered by the initiation of a tumble motion (and thus the CW rotation of a flagellum). In addition, tumbling times in the viscous fluid are similar to "bi-polar" swimming in the LC, supporting the hypothesis that the duration of a flagella being relocated on the other side of the bacterial body would essentially corresponds to the duration of a tumble.

This holds despite that fact that the unbundling and bundling mechanisms seem to be different in the two media, as we will discuss below. 
The observed tumble durations, defined as the duration during which the bundle is not fully formed, are larger compared to those in buffer solution, in agreement with the increased time needed to reform the bundle in a viscous surrounding. 
These observations comfort the idea that, in the LC case, the run time equals the duration of swimming in the "normal" configuration, whereas the tumble duration equals swimming in the bi-polar configuration. 
In the light of these observations we will refer to the reversal motion in the LCs as a frustrated run and tumble motion. 
In the next section, we will analyze in detail how a flagellum leaves the bundle to relocalize on the other side of the bacterial body. 

\subsection{Flagella relocation during  a frustrated tumble} 

\begin{figure*}[h]
\centering
\includegraphics[width=\textwidth]{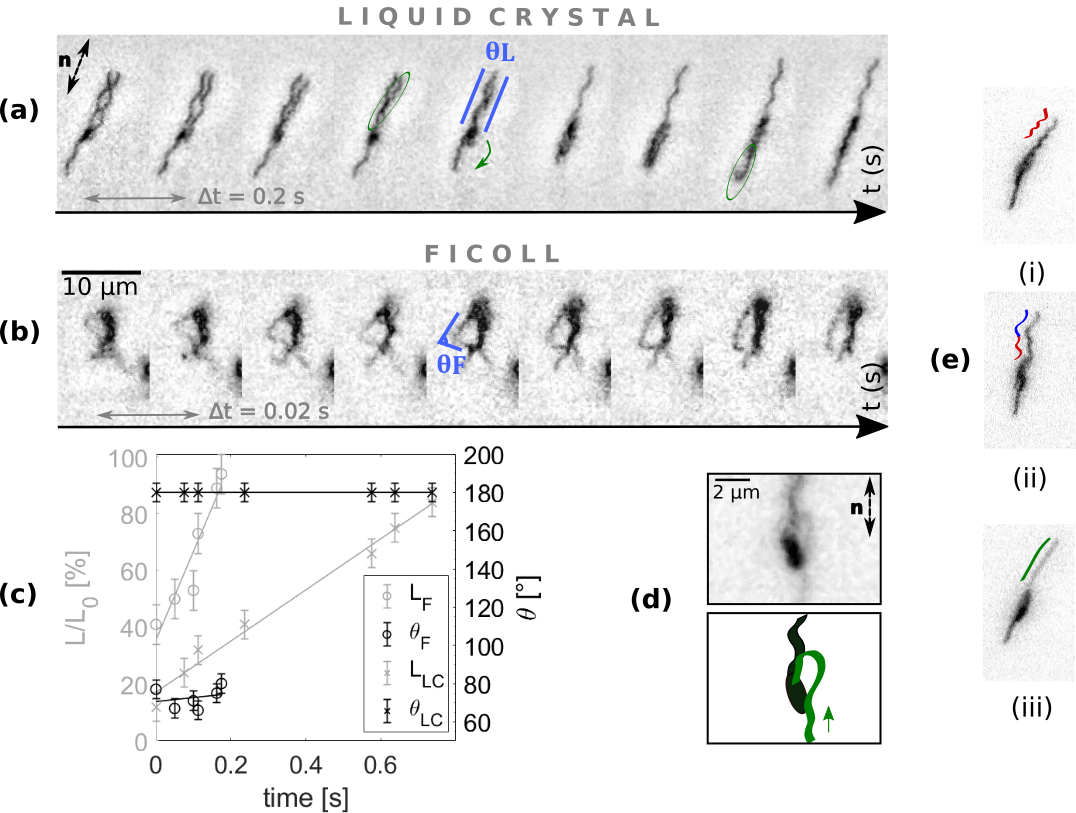}
\caption{(a) Flagellar relocation during a reversal in LC, accompanied by a polymorphic transformation. $\theta_L$ is the projected angle between the relocated flagellum (relocated by the green ellipse) and the bundle, here parallel (b) Snapshots of a polymorphic transformation occurring during a tumble, from normal to curly for a free swimming \textit{E. coli} bacterium in a viscous and Newtonian solution Ficoll. As indicated by the blue line, $\theta_F$ is the projected angle between a normal shaped portion of the flagellum and a curly portion. (c) Length from the body base to the deformation and kink projected angle $\theta$ over time, for a bacterium in Ficoll and one in DSCG liquid crystal (d) Zoom on the local deformation of the flagellum (in green). The arrow indicates the deformation propagation direction, travelling at a speed of $7.3 \pm 1 \si{\mu m.s^{-1}}$, and the double arrow the LC director. (e) Photographs of the same bacterium, presenting the zoology of flagellar shapes, and their illustrations : (i) curly II, in red (ii) unidentified curly in red - normal in blue (iii) unidentified straight shape, in green.}
\label{fig:kink-deformation}
\end{figure*}

We have seen that, in the LC, the flagella bundle, as well as the relocated flagellum, and the bacterial body, always remain aligned with the director field. This is shown in more detail in Fig. \ref{fig:kink-deformation}~(a), depicting the configurational transformations of a bacterium during the reversal phase in the LC. From these snapshots, we can see more clearly how the flagella relocation takes place, here during a transformation from a "bi-polar" to a "normal" configuration. The single flagellum folds along the bacterial body and then straightens out again on the other side to merge with the full bundle. This process involves rather strong deformations of the flagellar filament structure, as it can be seen in more detail in Fig. \ref{fig:kink-deformation}~(d), where the relocating flagellum adopts a hair pin configuration with a strong and localized curvature at the bending location.
We will first focus on the alignment with the director and then discuss a possible mechanism on how a flagellum is able to undergo such a strong deformation during relocation. 

Alignment of the bacterial body and the flagella bundle with the swimming direction induced by a surrounding complex media has already been observed, even if only indirectly, in elastic fluids, colloidal suspensions or colloidal liquid crystals \cite{Patteson2015d, Kamdar2022TheMotility,Lachlan2021}. Typically a misalignment between the flagella bundle and the bacterial body leads to wobbling of the body, easily observed under microscope \cite{Bianchi2017}. As a consequence, the reduction of wobbling amplitude and frequency has been used to conclude on the body-bundle alignment. Several studies, this time performed in simple Newtonian fluids, have linked decreased wobbling amplitudes, as well as more closed bundle forms, to more efficient swimming \cite{Bianchi2017, Najafi2018,Najafi2019}.  The better alignment is thus expected to be at the origin of the increased swimming speeds observed in complex fluids \cite{Patteson2015d, Kamdar2022TheMotility,Lachlan2021}.

It is thus not surprising that the elasticity of the LCs leads to bacteria-bundle alignment. Our observations constitute however for the first time direct visual evidence of this effect.  
However a completely new observation is the passage of the flagellum to the other side of the body involving the strong deformations described above (see Fig. \ref{fig:kink-deformation}~(d)). While no definitive explanation of the full mechanism of flagella reorientation can be given yet, we will in the following discuss some possible scenarios. 

First we will discuss whether the elasticity of the liquid crystal DSCG used in our study is important enough to sustain the strong deformations induced in the bacteria flagella. Bacterial flagella are know to be rather rigid with a bending stiffness of the order of $B \sim 3.5\ 10^{-24}\ \si{N m^2}$ \cite{Darnton2007}. An estimate of the elastic energy $E_{bend}$ stored when deforming a flagellum as in Fig. \ref{fig:kink-deformation}~(d) can be obtained taking only the localized deformation in the kink into account and by estimating the radius of curvature of the hairpin $R$ to be around 1 $\si{\mu m}$. On finds $E_{bend}\sim B/R \sim 10^{-18}Nm$. This can be compared to the elastic energy stored when deforming the liquid crystal director over a distance of the same order of magnitude as the radius of curvature. For the liquid crystal employed, in the one elastic constant approximation, at our experimental conditions, the elastic constant $K$ has a value   $K\sim 10 pN$ \cite{Zhou2014a}. The elastic energy associated to the director distortions induced by the bending of the flagellum can be estimated as $E_{elast} \sim KR \sim 10^{-17} \si{Nm}$. These two energies are comparable in magnitude. Another approach is to estimate the flagella deformation at which the sum of the elastic and bending energies is minimal, leading to a distance of $R\sim 1 \si{\mu m}$, in agreement with our observations. The liquid crystal thus seems able to sustain the strong deformation of the bent flagellum. 

The passage of the flagellum from one side of the body to the other is likely facilitated by a second mechanism: a polymorphic transformation. It is well known that polymorphic transformations can take place when the rotation direction of one flagellum in the bundle is changed from CCW to CW. Typically during such a transformation a normal shaped flagellum changes towards a so called "curly" shape, of different radius, pitch and chirality \cite{Darnton2007}. This transformation is initiated close to the bacterial body and then travels towards the free end of the flagellum. Geometrical constraints do not allow the connection between two different flagellar shapes to be straight at their junction. A natural kink, not involving any elastic deformation, necessary appears where the curly and normal flagellar segments meet. Such a kink travels along the flagellum as the polymorphic transformation advances. This has been observed for tethered cells by Darnton \etal \cite{Darnton2007}. 

Different types of flagellar shapes, with well defined geometries, have been predicted by \cite{Calladine1976}, and several experimentally observed, namely normal, curly I, curly II and semicoiled \cite{Darnton2007,Berg2005}. Curly shapes are characterized by a smaller pitch and thus a more "compact" visual appearance. The main difference between curly I and II lies in the total flagella length. Using this classification, we have measured the values of the pitch and radius of the helices for the normal and curly shapes in our experiments. The normal shape is characterized by a pitch and radius of respectively $3.37 \pm 0.20\ \si{\mu m}$ and $0.47 \pm 0.20\ \si{\mu m }$, while the curly shape has a pitch and radius of $1.47 \pm 0.20\ \si{\mu m}$ and $0.25\pm 0.20\ \si{\mu m}$, respectively. These pitch values are slightly larger than those reported in the literature for normal shaped \textit{E. coli} flagella, with a pitch of 2.3 \m ~and radius of 0.4 \m \cite{Berg2005}.
Generally, it is hard to distinguish between curly I and curly II shapes from the images, as their radius and pitch only differ by 0.1 \m, which is lower than the measurement error. However, we can conclude that we mostly observe curly II shapes, since all measurements indicate that the curly and normal shapes observed have the same length, $10 \pm 0.20\si{\mu m}$. Other shapes like semicoiled and curly 1 are usually the most observed in literature \cite{Darnton2007a, Turner2000, Darnton2007}. The different discrete flagella shapes are also associated with different handedness \cite{Calladine1976}. A curly flagellum can only exist as right handed, whereas a normal flagellum is left handed.

For the first time, an example of such a polymorphic transformation during a tumble of a freely swimming bacteria in the Ficoll solution is shown in Fig. \ref{fig:kink-deformation}~(b). Thanks to the high time resolution of the videos, the polymorphic transformation of the flagellum after its detachment from the bundle is clearly visible, as well as the angle $\theta$ between the normal-shaped portion of the flagellum (close to the free end) and the curly portion (close to the bacterial body). The fact that the shapes of the different filament morphologies are known with precision allows us to identify them unambiguously in our images. Close inspection of the flagella during their relocation in the LC (Fig. \ref{fig:kink-deformation}~(a)) reveals that also in this case a polymorphic transformation takes place. Indeed, one can observe two normal shaped flagella on the upper side of the body, until one of them, highlighted with a green ellipse, starts to change its shape to curly. This flagellum then migrates towards the bottom side of the body, while remaining aligned with the LC, parallel to the bacterial body. In the last snapshots, the bacterium is propelled upwards by the relocated flagellum, which remains in the curly shape. 

Fig. \ref{fig:kink-deformation}~(c) compares the angle of the kink during the polymorphic transformations shown in Figs \ref{fig:kink-deformation}~(a) and (b), for the LC and the Ficoll solution, respectively. In both cases, the angle remains nearly constant during the transformation and the kink propagates at nearly constant speed.   Note that the bacterium in Ficoll is freely swimming in space, and that the measured angle is a 2D projection. In the LC, the distinct flagellar segments are always parallel to the rest of the bacterium, resulting in a very small angle and giving the impression of sliding. The fact that this angle is significantly smaller compared to the observations in Ficoll might result from the combined action of the morphological transition and the elastic confinement the LC exerts onto the flagella segments. The deformation kink propagates at a speed of $v = 7.3 \pm 1\ \si{\mu m .s^{-1}}$ in the LC, and at a speed of $v = 20 \pm 1\ \si{\mu m .s^{-1}}$ in Ficoll. No literature values of the propagation speed of a polymorophic transformation under free swimming conditions exist and we can thus not compare our results to previous works. We have not performed a statistical analysis of these transformations because or their rarity in the field of view (in the case of Ficoll), but the examples shown represent the first quantitative measurement of the transformation dynamics. Despite the difference in absolute values, the two speeds observed are comparable in order of magnitude and point towards the fact that similar processes could be at play in both fluids. 

In some cases the flagella relocation in the LC can take place through more complex mechanisms and transformations. One example is shown in Fig. \ref{fig:kink-deformation}(e). Initially, the flagellum is in a curly I shape (Fig. \ref{fig:kink-deformation}(e)-(i)). Then, it transits to a hybrid configuration where two different shapes coexist (Fig. \ref{fig:kink-deformation}(e)-(ii)): one can see in image (ii) that the segment colored in red has a smaller pitch and radius than the segment colored in blue, the red segment corresponding to a curly shape and the blue segment to a normal shape. As expected, a finite angle is formed between the two distinct flagellar segments. This hybrid state remains stable during an extended period of time. We have rarely observed another shape than curly or normal, but in the image (iii), one can see an almost straight flagellum, a shape known to be unstable and never observed previously. 

The observations presented in this section on the relocation of a flagellum from one side of the body to the other side have demonstrated that this relocation takes place while the bacterium flagella and body remain aligned with the director field. The flagellum seems to "slide" along the bacterial body, a process that involves strong structural deformations. A combination between a morphological transformation, typically triggering the propagation of a kink along one of the flagella, and the elastic bending induced by the LC, seem to be at the origin of the observed dynamics.

\subsection{Swimming mechanism}

From our observations on flagella relocation and reversal initiation, a coherent picture of the swimming mechanism of \textit{E. coli} in the LC emerges. In the normal run phase, the flagella bundle pushes the bacterial body. At the onset of a tumble, a flagellum leaves the bundle, a morphological transformation from a normal to a curly shape occurs and the flagellum relocates to the other side of the bacterial body, now pushing the bacterium in the opposite direction. In this section, we address the question whether this swimming scenario is consistent with the zero force and torque condition required for swimming at low Reynolds number \cite{Lauga2007}. Besides, we try to understand how the action of a single flagellum can overcome the action of the whole bundle, placed on the other side of the bacterial body, pushing it in the opposite direction. 

Let us first state that bacteria swimming with a flagella bundle cannot reverse swimming direction by simply changing the rotation direction of the flagella bundle. This is in contrast with polar bacteria \cite{Stocker2012}, where a change in the rotation direction of the single flagellum indeed leads to a change from pushing to pulling the bacterial body. In the case of multiflagellated bacteria, as is the case of \textit{E. coli}, first a coordinated change in rotation direction of all flagella would have to take place simultaneously, which is never the case, but in addition, for the left-handed \textit{E. coli}, hydrodynamic interactions between flagella lead to flagella bundling for CCW rotations, but to unbundling for CW rotation and propulsion would thus be lost upon a change in rotation direction \cite{Kim2003}. This is why in normal swimming conditions multiflagellated bacteria perform the already described tumble mechanism to change swimming direction. 

Let us now discuss whether the flagellum relocated on the other side of the bacterial body is able to push the bacterium from a hydrodynamic point of view. The thrust produced by the rotating flagellum is compensated by the viscous friction on the moving body, following the zero force condition. In order to push the bacterial body forward,  flagella have to be either left-handed and turning CCW, or right-handed and turning CW. During a normal run the left-handed flagella in the bundle indeed turn CCW to push the cell. The single flagellum leaves the flagella bundle after the initiation of a tumble, and therefore, after a change of rotation direction. For the bacterium to be propelled by the relocated flagellum turning CW, the later must be right-handed. Handedness cannot be determined from our observations, representing a 2D projection of the flagella, however we can identify its morphological configuration from its geometrical properties. We have found the relocated flagella to be curly in almost all cases, which means that it is right-handed \cite{Calladine1976}. We can thus conclude that a curly flagellum rotating CW can indeed push a bacterium. 

\begin{figure*}[h]
\centering
\includegraphics[width=\textwidth]{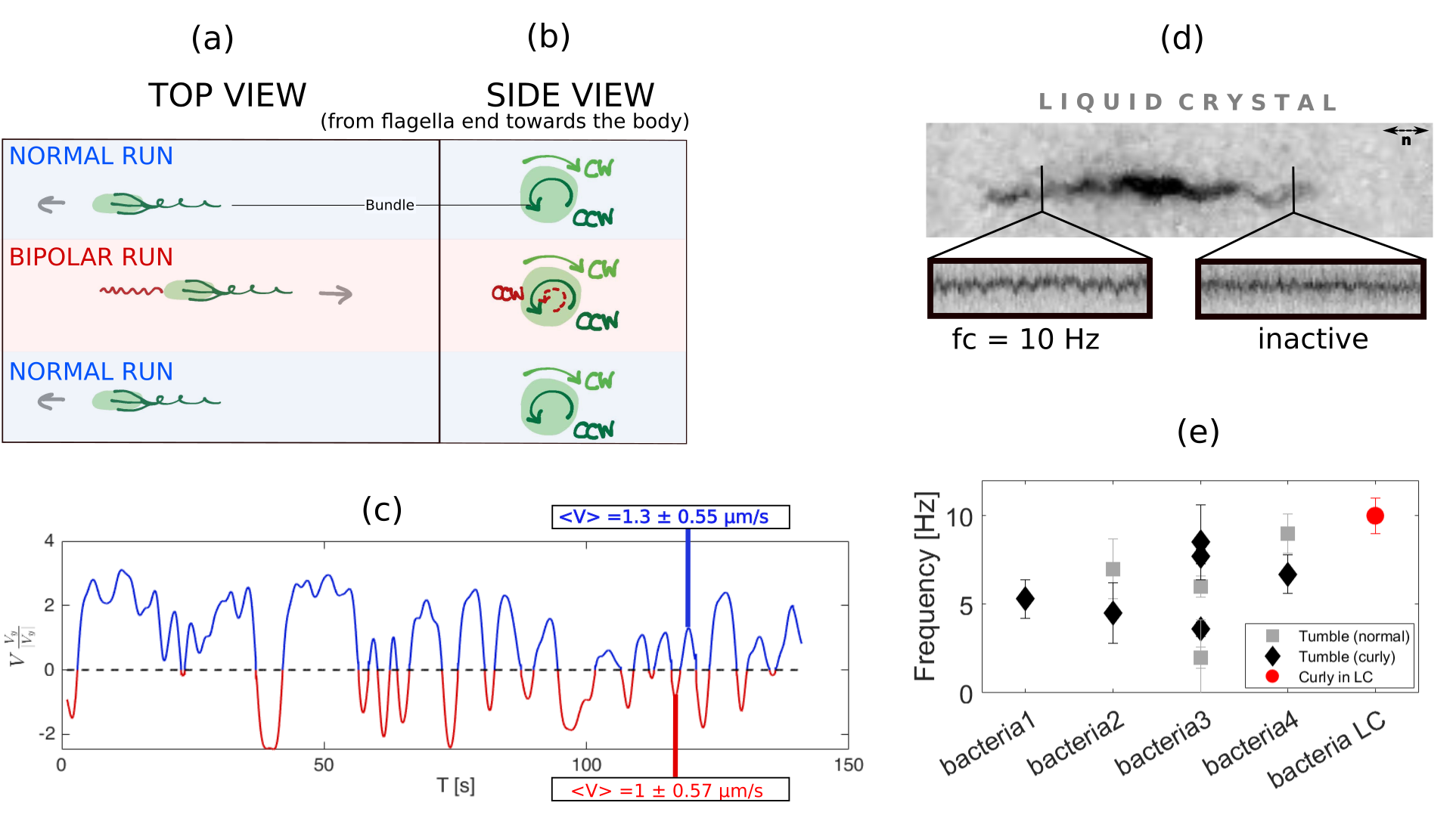}
\caption{(a) Illustration of the reorientation scenario for a bacterium swimming in a liquid crystal : first the bacterium is propelled by the bundle turning CCW in its reference frame, performing a normal run; in the red part, in a bipolar configuration, one flagellum changes its sense rotation to CW (and handedness) and migrates to the other side of the cell, it can be considered as a tumble. As the relocated flagellum is on the other side of the body, in the reference frame of the bundle, it turns CCW. The bacterium changed it swimming direction, it is propelled by the relocated flagellum. Finally the bacterium goes back to performing the normal run (blue part). (b) The side view is given for each step. (c) Swimming velocity $V_x$ projected along the forward swimming direction (prior to the first reversal) Each color corresponds to a different configuration (normal or bipolar). The general velocity averages for normal (blue) and bipolar (red) configurations (of all tracked bacteria) are $V = 1.3 \pm 0.55 \si{\mu m.s^{-1}}$ and $V = 1 \pm 0.57 \si{\mu m.s^{-1}}$ respectively. (d) Snapshot of a bacterium in LC, in bipolar configuration, and kymographs corresponding to the curly (of the left) and normal bundle (right) flagella. The bacterium swimming direction is from the left to the right, so that the curly flagellum is pushing it forward, by rotating at a 10 Hz frequency, while the bundle is inactive. (e)  Rotation frequencies for several bacteria swimming in Ficoll, during a tumble phase, for both curly and normal shapes flagella, comparable to the values of rotating curly flagellum in LC. }
\label{fig:mechanism_crosssection}
\end{figure*}

In addition, the zero torque condition also needs to be fulfilled for the bi-polar swimming configuration. During a normal run, all flagella turn in CCW direction and the body counter-rotates CW, maintaining zero torque. A schematic of the normal and bi-polar swimming configurations is represented in Fig. \ref{fig:mechanism_crosssection}(a), where we always observe from the direction of the flagella bundle towards the bacterial body. The flagellum leaving the bundle rotates CW, and therefore, in the same direction as the body. However once it has migrated to the other side of the body, from the perspective of the flagella bundle, it now rotates CCW, and thus, in the opposite direction to the bacterial body. Even if the details of the torques applied via the flagella hooks onto the bacterial body can be complex, from a geometrical point of view, the body and the pushing flagella counter-rotate again. Our swimming configuration on the bi-polar state is thus based on a coherent scenario in terms of sense of rotation of the body and flagella.
%It has been shown by Hintsche \etal \cite{Hintsche2017} that changing the motor sense of rotation is enough to propel the bacterium, enabling the swimming with a flagellum coiled around the body (for polar bacteria). 

In nearly all the cases where the flagellar configuration is bi-polar (around 70 bacteria tracks), we observe that the bacterium is pushed by the single flagellum against the flagella bundle. To understand the mechanism behind, we characterized the swimming speeds of the bacterium in the different configurations. We analysed the 64 exploitable tracks. From 36 tracks, with an average duration of $30 \pm 33\ \si{s}$, the bacterial trajectories did not show a reversal, while in 28 tracks, of $53 \pm 40\ \si{s}$ average duration, the bacteria performed reversals along their trajectory. For tracks without a reversal, the bacterium can either be in a normal or bipolar configuration for the whole duration of the experiment. We have used these trajectories to obtain clear information about the swimming speed in one or the other configuration. Reversals lead by definition to a slowing down of the swimming velocity, and therefore, they yield an overall reduction of the average velocity of the bacterium (see Fig. \ref{fig:trajectoire_vitesse_DSCG_Ficoll}~(c)). A bacterium swims in average with a velocity of $\overline{\rm v} = 4\pm 1\ \si{\mu m .s^{-1}}$, in the normal configuration, and $\overline{\rm v} = 2\pm0.5\ \si{\mu m .s^{-1}}$, in the bipolar configuration. Swimming in the bipolar configuration thus seems slower, but the difference is small. This is a surprising observation, as in the bi-polar configuration the single flagellum push against the flagella bundle. Even if, from a hydrodynamic point of view, the propulsion efficiency, at identical rotation rate and flagellar shape, is not higher for a flagella bundle than for a single flagellum, one would still expect the swimming speed in the normal configuration to be significantly larger.

 To gain more insight into this question, we have analyzed a typical example of a bacterium swimming in the bi-polar configuration in the LC, and we have measured the rotation frequencies of both the flagella bundle and the individual flagellum, see Fig. \ref{fig:mechanism_crosssection}(d). The frequency measurement has been done from kymographs along a defined line, perpendicular to the flagella. When the flagellum rotates, its wave propagates in space, thus the graph presents a wavy pattern. The abscissa being time, by measuring the separation between two peaks, one can get the frequency. Surprisingly, we find that the flagella bundle is nearly immobile and rotates with a frequency below $1 Hz$. The curly individual flagella on the other side of the bacterial body, in contrast, rotates at a frequency of nearly $10 Hz$. For comparison, we measured rotation frequencies for bacteria swimming in Ficoll. The results presented in Fig. \ref{fig:mechanism_crosssection}(e) are obtained during the tumble and show that the curly and normal flagella rotate at a similar frequencies, between $5$ and $10 Hz$, comparable to the observed rotation frequency of the single flagellum in the bi-polar configuration. Without being able to fully explain this observation, namely why the flagella bundle becomes nearly immobile, this is in agreement with a displacement of the bacterial body pushed by the single flagellum against a passive flagella bundle. It has been demonstrated that the swimming efficiency for curly shaped flagella is slightly smaller compared to normal flagella \cite{Spagnolie}, which could be among the reasons why the bi-polar swimming velocity is observed to be smaller than the normal swimming velocity. The proposed mechanism for bi-polar swimming obtained from our experimental observation is thus also consistent from a hydrodynamic point of view. 

\section{Conclusion}

In this report we described how wild-type \textit{E. coli} bacteria explore space in a structured fluid like a liquid crystal (LC).  In such a fluid the swimming direction is constrained to take place along the nematic director and we observed that bacterial re-orientation occurs in the form of 180°  reversals. This is in stark contrast with typical trajectories of wild-type bacteria exploring isotropic fluids that resemble a 3D random walk, stemming from the classical run and tumble dynamics.

Precise experimental observations of the swimming bacteria, including full monitoring of the flagellar dynamics, revealed that these reversals always occur together with a specific flagella rearrangement, implying a flagellum leaving the bundle and repositioning itself on the other side of the bacterial body. We believe that this remarkable relocation is triggered by a reverse in rotation direction of the repositioning flagellum, a mechanism that, in an isotropic fluid, would lead to a classical tumble, following the opening of the flagella bundle.

To support this hypothesis, we performed experiments in a Newtonian fluid of  viscosity comparable to that of the LC. Typical swimming times using the full bundle (here called normal configuration) are comparable to typical run times in the viscous fluid, while swimming times using the relocated single flagellum (here called bi-polar configuration) are comparable to typical tumble times. This suggests that the reversals stem from a frustrated tumble. 

Precise and time resolved observations were made possible by the slowing down of the bacterial dynamics in the viscous surroundings and an excellent resolution of the flagellar structure revealed a precise scenario for the repositioning dynamics. The repositioning of the flagellum on the other side of the bacterial body takes place,  while the bacterium flagella and body remain aligned along the director field. The flagellum seems to "slide" along the bacterial body, a process that involves strong structural deformations, in contrast to the typical opening of the flagella bundle observed in isotropic media. A combination between a morphological transformation, typically triggering the propagation of a kink along one of the bacterium flagella, and the elastic bending induced by the LC, seems to be at the origin of the observed flagellar dynamics.
Relocation of the single flagellum on the other side of the bacterial body coincides in nearly all cases with a change in swimming direction, meaning that the single flagellum pushes the bacterial body against the flagella bundle located on the opposite side. Close inspection of bundle and flagella rotation frequencies indicate that, in these cases, the flagella bundle becomes nearly immobile and thus passive to propulsion.   Meanwhile, the single flagella rotates at frequencies of comparable magnitude to those observed in the corresponding viscous fluid. While we have no clear explanation for the observed stalling of the flagella bundle, our observations are consistent with a swimming mechanism at low Reynolds number. 

We believe that our experimental characterization of the novel bacterial reversal dynamics observed in LCs will constitute the basis of future modeling and research. Some observations represent an extended characterization of representative examples instead of a full statistical analysis, but in most cases they constitute the first observation ever made of such dynamics, in the context of highly viscous Newtonian viscous fluid as well as structured LCs. As such, these observations could also pave the way to further experimental observations of comparable phenomena including bacteria flagella dynamics in various complex environments.

\begin{acknowledgments}
 We thank Prof. Dong Ki Yoon, from Korea Advanced Institute of Science and Technology (KAIS), who has kindly provided us with patterned glass slides for the alignment of the LC. We thank Prof Wilson Poon's group of the University of Edimburgh, for providing us with the two-colours  strain AD62. We also thank Anumita Jawahar for measuring the phase diagram presented in the Appendix. We thank Prof. Alexander Morozov for enlightening discussions on possible reversal and swimming mechanisms. We thank Prof. Kenny Breuer for bringing the possible role of morphological transformations to our attention and Guillaume Sintes for helpful discussions on flagella dynamics and bacteria swimming.
\end{acknowledgments}

\def\aucontribute#1{{\vskip5.5pt\noindent \textbf{{\fontsize{10}{11}\selectfont Authors' Contributions :}}\fontsize{8}{11}\selectfont\enskip #1}}
\aucontribute{MG, EC, TLL and AL have designed the study. MG has performed the experiments and analyzed the data, TD has developed and implemented the two color tracking technique. MG, TLL and AL have written the paper.}

\def\funding#1{{\vskip5.5pt\noindent \textbf{{\fontsize{10}{11}\selectfont Funding :}}\fontsize{8}{11}\selectfont\enskip #1}}
\funding{This work has been funded by the Agence Nationale de la Recherche (ANR) grant ANR-13-JS08-0006-01 and the European Research Council (ERC) Consolidator Grant ‘PaDyFlow’, Grant Agreement no. 682367. We acknowledge the Institut Pierre-Gilles de Gennes (equipement d’excellence, Investissements d’avenir program ANR-10-EQPX-34).}

%\vfill\eject
\appendix

\section{Appendixes}
\subsection{DSCG phase diagram}

\begin{figure*}[ht]
\centering
\includegraphics[width=8cm]{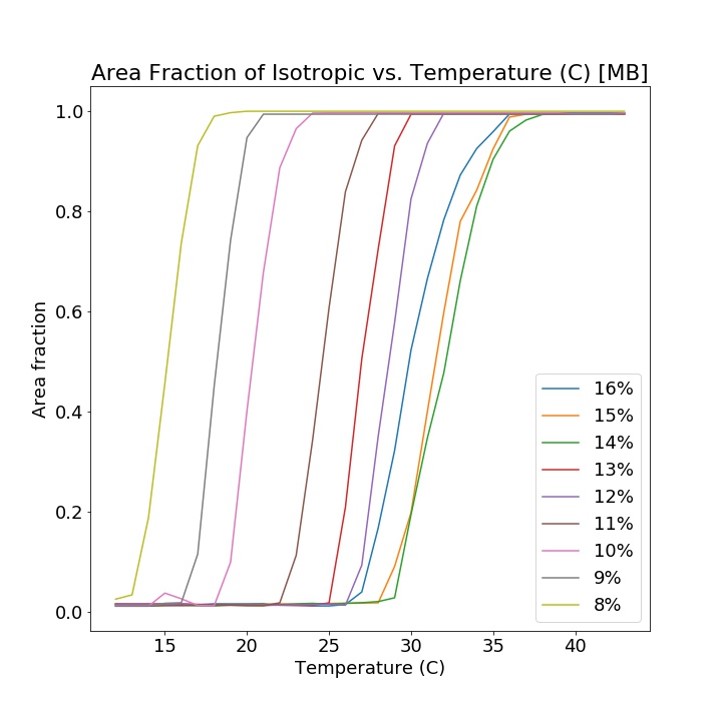}
\caption{Area fraction of DSCG (dispersed in MB) isotropic phase as a function of the temperature, for several concentrations.}
\label{diagram}
\end{figure*}

The DSCG liquid crystal (LC) was dispersed in a motility buffer, thus we established the phase diagram for this system, shown in the Figure \ref{diagram}.

\begin{figure*}[ht]
\centering
\includegraphics[width=\textwidth]{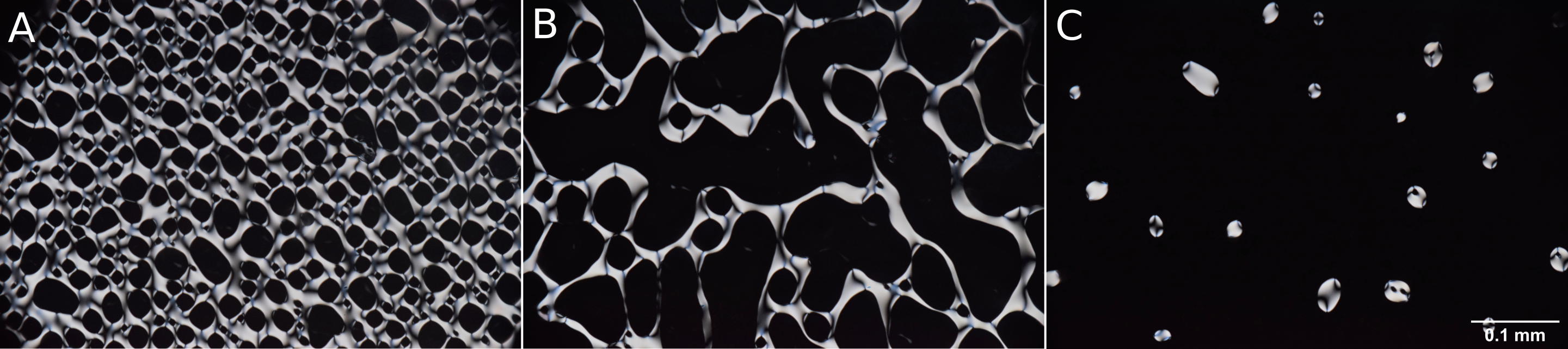}
\caption{Crossed-polarized images of the temperature-induced nematic to isotropic phase transition for 15.8 \% DSCG dissolved in pure water (A) 34°  C (B) 36°  C (C) 38°  C. In black the isotropic phase and in white the nematic phase.}
\label{tactoids}
\end{figure*}

The study consisted in varying the temperature and concentration of several LC mixtures. For each concentration (from 8 to 16 wt\%), the temperature was varied from 12°to 43° using a microscope TS64 thermal stage. The samples were sealed with nail polish to prevent any water evaporation that could induce an increase of the DSCG concentration. 
In order to ensure a thin layer of LC, no spacers were used. 
Using a custom Matlab code, 4 regions on the sample were selected for the FOV in order to increase the statistics. 
Images were taken using a cross-polarizer and after waiting for several minutes until the LC mixture relaxed. 
An example of the resulting images is shown in the Figure \ref{tactoids} with isotropic regions called tactoids (in black) getting bigger with increasing temperature. 
A custom Python code was written to analyze the images, detecting the isotropic and nematic phases on binarized images, in order to get the corresponding area fractions.
On the phase diagram, when the isotropic fraction equals one, the LC is in a fully isotropic phase. For a zero isotropic fraction the LC is in a fully nematic phase. In between, the DSCG has mixed regions of isotropic and nematic, like shown in the Figure \ref{tactoids}. In our case, we want to avoid tactoids because they disturb bacterial motion \cite{Zhou2014a}. 

\subsection{Experimental set-up for sample preparation}
\begin{figure*}[!ht]
\centering
\includegraphics[width=8cm]{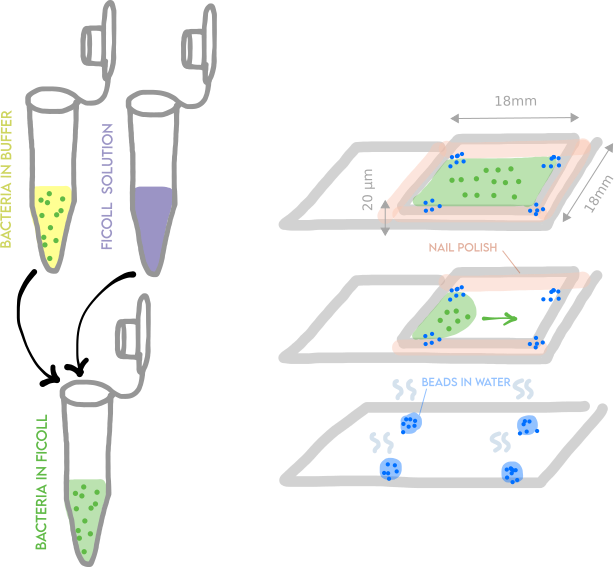}
\caption{Sample preparation for the study of bacterial swimming in aligned LC.}
\label{setup}
\end{figure*}

See Figure \ref{setup}.

%\nocite{*}
\newpage~\newpage
\bibliography{references.bib}

\end{document}